%% file: main.tex
\documentclass[letterpaper,twocolumn,10pt]{article}
\usepackage{usenix2019_v3}
\usepackage{tikz}

\microtypecontext{spacing=nonfrench}

\usepackage{times}

\usepackage{prp-macros}

\usepackage{listings}
\usepackage{makecell}
\usepackage{textcomp}
\usepackage{hyperref}
\usepackage{multirow}
\usepackage{amsfonts}

\usepackage{algorithm}
\usepackage[noend]{algpseudocode}

\usepackage{caption}
\usepackage[small,compact]{titlesec}

\usepackage{subcaption}

\hypersetup{
    colorlinks=true,
    linkcolor=blue,
    filecolor=orange,      
    urlcolor=blue,
}

\definecolor{dkgreen}{rgb}{0,0.6,0}
\definecolor{gray}{rgb}{0.5,0.5,0.5}
\definecolor{dkgray}{rgb}{0.3,0.3,0.3}
\definecolor{mauve}{rgb}{0.58,0,0.82}
\definecolor{dkorange}{rgb}{0.7,0.1,0}

\usepackage{inconsolata}

\newcommand{\Faaslets}{Faaslets\xspace}
\newcommand{\Faaslet}{Faaslet\xspace}
\newcommand{\sys}{\textsc{Faasm}\xspace}

\begin{document}

\title{\sys: Lightweight Isolation for Efficient Stateful Serverless Computing}

\author{
{\rm Simon Shillaker}\\
Imperial College London
\and
{\rm Peter Pietzuch}\\
Imperial College London
}

\maketitle

\begin{abstract}
  Serverless computing is an excellent fit for big data processing because it
  can scale quickly and cheaply to thousands of parallel functions. Existing
  serverless platforms isolate functions in ephemeral, stateless containers,
  preventing them from directly sharing memory. This forces users to duplicate
  and serialise data repeatedly, adding unnecessary performance and resource
  costs. We believe that a new lightweight isolation approach is needed, which
  supports sharing memory directly between functions and reduces resource
  overheads.
  
  We introduce \emph{\Faaslets}, a new isolation abstraction for
  high-performance serverless computing. \Faaslets isolate the memory of
  executed functions using \emph{software-fault isolation}~(SFI), as provided by
  \emph{WebAssembly}, while allowing memory regions to be shared between
  functions in the same address space. \Faaslets can thus avoid expensive data
  movement when functions are co-located on the same machine. Our runtime for
  \Faaslets, \sys, isolates other resources, \eg CPU and network, using standard
  Linux \emph{cgroups}, and provides a low-level POSIX host interface for
  networking, file system access and dynamic loading. To reduce initialisation
  times, \sys restores \Faaslets from already-initialised snapshots. We compare
  \sys to a standard container-based platform and show that, when training a
  machine learning model, it achieves a 2$\times$ speed-up with 10$\times$ less
  memory; for serving machine learning inference, \sys doubles the throughput
  and reduces tail latency by 90\%.
\end{abstract}

\input{sections/intro}

\input{sections/background}

\input{sections/faaslets}

\input{sections/state}

\input{sections/arch}

\input{sections/evaluation}

\input{sections/related}

\input{sections/conclusion}

\input{sections/acks}

{\footnotesize\bibliographystyle{plain}
\bibliography{main}}

\end{document}

%% file: sections/intro.tex
% Introduction

\section{Introduction}
\label{sec:intro}

% Serverless is a good fit...
Serverless computing is becoming a popular way to deploy
data-intensive applications. A function-as-a-service~(FaaS) model decomposes
computation into many functions, which can effectively exploit the massive
parallelism of clouds. Prior work has shown how serverless can support
map/reduce-style jobs~\cite{PyWren:2017, IBMPywren:2018}, machine learning
training~\cite{ServerlessML:2018, Cirrus:2019} and
inference~\cite{ishakian2018serving}, and linear algebra
computation~\cite{werner2018serverless, Numpywren:2018}. As a result, an
increasing number of applications, implemented in diverse programming languages,
are being migrated to serverless platforms.

%... but external state and containers aren't
Existing platforms such as Google Cloud Functions~\cite{GCFWebsite}, IBM Cloud
Functions~\cite{IBMCloudFunctionsWebsite}, Azure
Functions~\cite{AzureFunctionsWebsite} and AWS Lambda~\cite{AWSLambdaWebsite}
isolate functions in ephemeral, stateless \emph{containers}. The use of
containers as an isolation mechanisms introduces two challenges for
data-intensive applications, \emph{data access overheads} and the
\emph{container resource footprint}. 

% Data overheads
Data access overheads are caused by
the stateless nature of the container-based approach, which forces state to be
maintained externally, \eg in object stores such as
Amazon~S3~\cite{AWSS3Website}, or passed between function invocations. Both
options incur costs due to duplicating data in each function, repeated
serialisation, and regular network transfers. This results in current
applications adopting an inefficient ``data-shipping architecture'', \ie moving
data to the computation and not vice versa---such architectures have been
abandoned by the data management community many decades
ago~\cite{OneStepForward:2018}. These overheads are compounded as the number of
functions increases, reducing the benefit of unlimited parallelism, which is
what makes serverless computing attractive in the first place.

% Container overheads
The container resource footprint is particularly relevant because of the
high-volume and short-lived nature of serverless workloads. Despite containers having a smaller memory and CPU overhead than
other mechanisms such as virtual machines~(VMs), there remains an impedance
mismatch between the execution of individual short-running functions and the
process-based isolation of containers. Containers have start-up latencies in the
hundreds of milliseconds to several seconds, leading to the \emph{cold-start}
problem in today's serverless platforms~\cite{Peeking:2018,
OneStepForward:2018}. The large memory footprint of containers limits
scalability---while technically capped at the process limit of a machine, the
maximum number of containers is usually limited by the amount of available
memory, with only a few thousand containers supported on a machine with
16\unit{GB} of RAM~\cite{LightVM}.

% Current solutions and workarounds

Current data-intensive serverless applications have addressed these
problems individually, but never solved both---instead, either exacerbating the container resource overhead or breaking the serverless model.
Some systems avoid data movement costs by
maintaining state in long-lived VMs or services, such as
ExCamera~\cite{ExCamera:2017}, Shredder~\cite{Shredder:2019} and
Cirrus~\cite{Cirrus:2019}, thus introducing non-serverless components. To 
address the performance overhead of
containers, systems typically increase the level of trust in users' code
and weaken isolation guarantees.
PyWren~\cite{PyWren:2017} reuses containers to execute multiple functions;
Crucial~\cite{Crucial:2019} shares a single instance of the Java virtual
machine~(JVM) between functions; SAND~\cite{SAND:2018} executes multiple
functions in long-lived containers, which also run an additional message-passing
service; and Cloudburst~\cite{Cloudburst:2020} takes a similar approach,
introducing a local key-value-store cache. Provisioning containers to execute multiple functions and extra
services amplifies resource overheads, and breaks the fine-grained elastic
scaling inherent to serverless. While several of these systems reduce data
access overheads with local storage, none provide \emph{shared memory} between
functions, thus still requiring duplication of data in separate process
memories.

% Existing work on isolation mechanisms
Other systems reduce the container resource footprint
by moving away from containers and VMs. Terrarium~\cite{FastlyTerrariumWebsite}
and Cloudflare Workers~\cite{CloudflareWorkersWebsite} employ software-based
isolation using WebAssembly and V8 Isolates, respectively;
Krustlet~\cite{KrustletWebsite} replicates containers using WebAssembly for
memory safety; and SEUSS~\cite{SEUSS:2020} demonstrates serverless unikernels.
While these approaches have a reduced resource footprint, they do not address data
access overheads, and the use of software-based isolation alone does not isolate
resources.

% Why is this hard? Tension between isolation and state
We make the observation that serverless computing can better support
data-intensive applications with a new isolation abstraction that (i)~provides strong memory and resource isolation between
functions, yet (ii)~supports efficient state sharing. Data should be
\emph{co-located} with functions and accessed directly, minimising
data-shipping. Furthermore, this new isolation abstraction must (iii)~allow
scaling state across multiple hosts; (iv)~have a low memory footprint,
permitting many instances on one machine; (v)~exhibit fast instantiation times;
and (vi)~support multiple programming languages to facilitate the porting of
existing applications.

% Contributions
In this paper, we describe \textbf{\Faaslets}, a new \emph{lightweight
  isolation abstraction} for data-intensive serverless computing. \Faaslets
support stateful functions with efficient shared memory access, and are
executed by our \textbf{\sys} distributed serverless runtime. \Faaslets have
the following properties, summarising our contributions:

% Isolation with shared memory
\mypar{(1)~\Faaslets achieve lightweight isolation} \Faaslets rely on
\emph{software fault isolation}~(SFI)~\cite{Wahbe:1993}, which restricts
functions to accesses of their own memory. A function associated with a
\Faaslet, together with its library and language runtime dependencies, is
compiled to WebAssembly~\cite{wasm2017}. The \sys runtime then executes
multiple \Faaslets, each with a dedicated thread, within a single address
space. For resource isolation, the CPU cycles of each thread are constrained
using Linux \emph{cgroups}~\cite{LinuxKernelWebsite} and network access is
limited using \emph{network namespaces}~\cite{LinuxKernelWebsite} and
\emph{traffic shaping}.  Many \Faaslets can be executed efficiently and safely
on a single machine.

% Shared state
\mypar{(2)~\Faaslets support efficient local/global state access} Since
\Faaslets share the same address space, they can access shared memory regions
with local state efficiently. This allows the co-location of data and functions
and avoids serialisation overheads. Faaslets use a two-tier state architecture,
a \emph{local} tier provides in-memory sharing, and a \emph{global} tier
supports distributed access to state across hosts. The \sys runtime provides a
state management API to \Faaslets that gives fine-grained control over state
in both tiers. \Faaslets also support stateful applications with different
consistency requirements between the two tiers.

% Proto-Faaslets
\mypar{(3)~\Faaslets have fast initialisation times} To reduce cold-start time
when a \Faaslet executes for the first time, it is launched from a suspended
state. The \sys runtime pre-initialises a Faaslet ahead-of-time and snapshots
its memory to obtain a \emph{Proto-\Faaslet}, which can be restored in
hundreds of microseconds. Proto-\Faaslets are used to create fresh \Faaslet
instances quickly, \eg avoiding the time to initialise a language runtime.
While existing work on snapshots for serverless takes a
single-machine approach~\cite{SAND:2018,SOCK:2018,Catalyzer:2020,SEUSS:2020},
Proto-Faaslets support cross-host restores and are OS-independent.

% Host interface/ dynamic language support
\mypar{(4)~\Faaslets support a flexible host interface} \Faaslets interact with
the host environment through a set of POSIX-like calls for networking, file I/O,
global state access and library loading/linking. This allows them to support
dynamic language runtimes and facilitates the porting of existing applications,
such as CPython by changing fewer than 10 lines of code. The host
interface provides just enough virtualisation to ensure isolation while adding a
negligible overhead.

\tinyskip

% Pluggability

\noindent
The \sys runtime\footnote{\sys is open-source and available 
at \href{https://github.com/lsds/Faasm} {github.com/lsds/Faasm}} uses the LLVM 
compiler toolchain to translate applications to
WebAssembly and supports functions written in a range of programming languages,
including C/C++, Python, Typescript and Javascript. It integrates with existing
serverless platforms, and we describe the use with
\emph{Knative}~\cite{KNativeGithub}, a state-of-the-art platform based on
Kubernetes.

% Evidence/experiments
To evaluate \sys{}'s performance, we consider a number of workloads and compare
to a container-based serverless deployment. When training a machine learning
model with SGD~\cite{Hogwild:2011}, we show that \sys achieves a 60\%
improvement in run time, a 70\% reduction in network transfers, and a 90\%
reduction in memory usage; for machine learning inference using TensorFlow
Lite~\cite{TFLiteWebsite} and MobileNet~\cite{MobileNets:2017}, \sys achieves
over a 200\% increase in maximum throughput, and a 90\% reduction in tail
latency. We also show that \sys executes a distributed linear algebra job for
matrix multiplication using Python/Numpy with negligible performance overhead
and a 13\% reduction in network transfers.

% End

%% file: sections/background.tex
% Background/ Motivation

\section{Isolation vs. Sharing in Serverless}
\label{sec:background}

Sharing memory is fundamentally at odds with the goal of isolation, hence
providing shared access to in-memory state in a multi-tenant serverless
environment is a challenge. 

% Functional
\T\ref{table:Alternatives} contrasts \emph{containers} and \emph{VMs} with other
potential serverless isolation options, namely
\emph{unikernels}~\cite{SEUSS:2020} in which minimal VM images are used to pack
tasks densely on a hypervisor and \emph{software-fault
isolation}~(SFI)~\cite{Wahbe:1993}, providing lightweight memory safety through
static analysis, instrumentation and runtime traps. The table lists whether each
fulfils three key functional requirements: memory safety, resource isolation and
sharing of in-memory state. A fourth requirement is the ability to share a
filesystem between functions, which is important for legacy code and to reduce
duplication with shared files.

% Non-functional
The table also compares these options on a set of non-functional requirements:
low initialisation time for fast elasticity; small memory footprint for
scalability and efficiency, and the support for a range of programming
languages.

% Comparison and drawbacks summary
Containers offer an acceptable balance of features if one sacrifices efficient
state sharing---as such they are used by many serverless
platforms~\cite{GCFWebsite, IBMCloudFunctionsWebsite,
  AzureFunctionsWebsite}. Amazon uses Firecracker~\cite{FirecrackerGithub}, a
``micro VM'' based on KVM with similar properties to containers, \eg
initialisation times in the hundreds of milliseconds and memory overheads of
megabytes.

% Comparison of alternatives
\begin{table}[t]
    \centering\footnotesize
    \begin{tabular}{ll@{}c@{}c@{}c@{}c@{}c}
    \toprule
        &  & \textbf{Containers} & \textbf{VMs} & \textbf{Unikernel} & \textbf{SFI} & \textbf{Faaslet} \\    
    \midrule
        \multirow{4}{*}{\rotatebox[origin=c]{90}{\scriptsize{}Func.}} & Memory safety & \cmark & \cmark & \cmark & \cmark & \cmark\\
    % \cline{2-7}
        & Resource isolation & \cmark & \cmark & \cmark & \xmark & \cmark \\    
    % \cline{2-7}
        & Efficient state sharing & \xmark & \xmark & \xmark & \xmark & \cmark \\
    % \cline{2-7}
        & Shared filesystem & \cmark & \xmark & \xmark & \cmark & \cmark \\
      \midrule
      \multirow{3}{*}{\rotatebox[origin=c]{90}{\parbox{0.5cm}{\scriptsize{}Non-func.}}} & Initialisation time & \SI{100}{\ms}  & \SI{100}{\ms} & \SI{10}{\ms} & \SI{10}{\micro\second}& \SI{1}{\ms} \\
    % \cline{2-7}
        & Memory footprint & MBs & MBs & KBs & Bytes & KBs \\
    % \cline{2-7}
        & Multi-language & \cmark & \cmark & \cmark & \xmark & \cmark \\
    \bottomrule
    \end{tabular}
    \caption{Isolation approaches for serverless {\textnormal{(Initialisation times include ahead-of-time snapshot restore where applicable~\cite{SOCK:2018,Catalyzer:2020,SEUSS:2020}.)}}}
    \label{table:Alternatives}
\end{table}

Containers and VMs compare poorly to unikernels and SFI on initialisation times
and memory footprint because of their level of virtualisation. They both
provide complete virtualised POSIX environments, and VMs also virtualise
hardware. Unikernels minimise their levels of virtualisation, while SFI
provides none. Many unikernel implementations, however, lack the maturity
required for production serverless platforms, \eg missing the required tooling
and a way for non-expert users to deploy custom images. SFI alone
cannot provide resource isolation, as it purely focuses on memory safety. It
also does not define a way to perform isolated interactions with the underlying
host. Crucially, as with containers and VMs, neither unikernels nor SFI can
share state efficiently, with no way to express shared memory regions between
compartments.

\subsection{Improving on Containers}

% Summary of the problem and what's been done
Serverless functions in containers typically share state via external storage
and duplicate data across function instances. Data access and serialisation
introduces network and compute overheads; duplication bloats the memory
footprint of containers, already of the order of
megabytes~\cite{LightVM}. Containers contribute hundreds of milliseconds up to
seconds in cold-start latencies~\cite{Peeking:2018}, incurred on initial
requests and when scaling. Existing work has tried to mitigate these drawbacks
by recycling containers between functions, introducing static VMs, reducing
storage latency, and optimising initialisation.

% Recycling containers, sacrificing isolation
Recycling containers avoids initialisation overheads and allows data caching but
sacrifices isolation and multi-tenancy. PyWren~\cite{PyWren:2017} and its
descendants, Numpywren~\cite{Numpywren:2018}, IBMPywren~\cite{IBMPywren:2018},
and Locus~\cite{Locus:2019} use recycled containers, with long-lived AWS Lambda
functions that dynamically load and execute Python functions.
Crucial~\cite{Crucial:2019} takes a similar approach, running multiple functions
in the same JVM. SAND~\cite{SAND:2018} and Cloudburst~\cite{Cloudburst:2020}
provide only process isolation between functions of the same application and
place them in shared long-running containers, with at least one additional
background storage process. Using containers for multiple functions and
supplementary long-running services requires over-provisioned memory to ensure
capacity both for concurrent executions and for peak usage. This is at odds with
the idea of fine-grained scaling in serverless.

% Static storage, breaking serverless
Adding static VMs to handle external storage improves performance but breaks
the serverless paradigm. Cirrus~\cite{Cirrus:2019} uses large VM instances to
run a custom storage back-end; Shredder~\cite{Shredder:2019} uses a single
long-running VM for both storage and function execution;
ExCamera~\cite{ExCamera:2017} uses long-running VMs to coordinate a pool of
functions. Either the user or provider must scale these VMs to match the
elasticity and parallelism of functions, which adds complexity and
cost.

% Optimized storage, helping latency but not data shipping
Reducing the latency of auto-scaled storage can improve performance within the
serverless paradigm. Pocket~\cite{Pocket:2018} provides ephemeral serverless
storage; other cloud providers offer managed external state, such as AWS Step
Functions~\cite{AWSStepWebsite}, Azure Durable
Functions~\cite{AzureDurableWebsite}, and IBM
Composer~\cite{OpenwhiskComposerWebsite}. Such approaches, however, do not
address the data-shipping problem and its associated network and memory
overheads.

% Optimizing containers, diminishing returns
Container initialisation times have been reduced to mitigate the cold-start
problem, which can contribute several seconds of latency with standard
containers~\cite{ServerlessArchitectrual:2019, OneStepForward:2018,
Peeking:2018}. SOCK~\cite{SOCK:2018} improves the container boot process to
achieve cold starts in the low hundreds of milliseconds;
Catalyzer~\cite{Catalyzer:2020} and SEUSS\cite{SEUSS:2020} demonstrate
snapshot and restore in VMs and unikernels to achieve millisecond serverless
cold starts. Although such reductions are promising, the resource overhead and
restrictions on sharing memory in the underlying mechanisms still remain.

\subsection{Potential of Software-based Isolation}

Software-based isolation offers memory safety with initialisation times
and memory overheads up to two orders of magnitude lower than containers and
VMs. For this reason, it is an attractive starting point for serverless
isolation. However, software-based isolation alone does not support resource 
isolation, or efficient in-memory state sharing.

It has been used in several existing edge and serverless computing systems, but
none address these shortcomings. Fastly's Terrarium~\cite{FastlyTerrariumWebsite}
and Cloudflare Workers~\cite{CloudflareWorkersWebsite} provide memory safety
with WebAssembly~\cite{wasm2017} and V8 Isolates~\cite{V8Github}, respectively,
but neither isolates CPU or network use, and both rely on data shipping for
state access; Shredder~\cite{Shredder:2019} also uses V8 Isolates to run code on
a storage server, but does not address resource isolation, and relies on
co-locating state and functions on a single host. This makes it ill-suited to
the level of scale required in serverless platforms;
Boucher~\etal\cite{Boucher:2018} show microsecond initialisation times for Rust
microservices, but do not address isolation or state sharing;
Krustlet~\cite{KrustletWebsite} is a recent prototype using WebAssembly to
replace Docker in Kubernetes, which could be integrated with
Knative~\cite{KNativeGithub}. It focuses, however, on replicating
container-based isolation, and so fails to meet our requirement for in-memory
sharing.

Our final non-functional requirement is for multi-language support, which
is not met by language-specific approaches to software-based
isolation~\cite{RustSystems:2017,SingularityMessage:2006}. Portable Native
Client~\cite{PNacl:2010} provides multi-language software-based isolation by
targeting a portable intermediate representation, LLVM IR, and hence meets this
requirement. Portable Native Client has now been deprecated, with WebAssembly as
its successor~\cite{wasm2017}.

WebAssembly offers strong memory safety guarantees by constraining memory access
to a single linear byte array, referenced with offsets from zero. This enables
efficient bounds checking at both compile- and runtime, with runtime checks
backed by traps. These traps (and others for referencing invalid functions) are
implemented as part of WebAssembly runtimes~\cite{WasmSpec}. The security
guarantees of WebAssembly are well established in existing literature, which
covers formal verification~\cite{MechanisingWasm:2018}, taint
tracking~\cite{TaintAssembly:2018}, and dynamic analysis~\cite{Wasabi:2019}. WebAssembly offers mature support for
languages with an LLVM front-end such as C, C++, C\#, Go and
Rust~\cite{LLVM9Website}, while toolchains exist for
Typescript~\cite{AssemblyScriptGithub} and Swift~\cite{SwiftWasmWebsite}. Java
bytecode can also be converted~\cite{TeaVMWebsite}, and further language support
is possible by compiling language runtimes to WebAssembly, \eg Python,
JavaScript and Ruby. Although WebAssembly is restricted to a 32-bit address
space, 64-bit support is in development.

The WebAssembly specification does not yet include mechanisms for
sharing memory, therefore it alone cannot meet our requirements. There is a
proposal to add a form of synchronised shared memory to
WebAssembly~\cite{WeakeningWasm:2019}, but it is not well suited to 
sharing serverless state dynamically due to the required compile-time knowledge of all shared
regions. It also lacks an associated programming model and provides only local
memory synchronisation.

The properties of software-based isolation highlight a compelling
alternative to containers, VMs and unikernels, but none of these approaches meet
all of our requirements. We therefore propose a new isolation approach to
enable efficient serverless computing for big data.

% End

%% file: sections/faaslets.tex
% Faaslets

\section{Faaslets} \label{sec:faaslet}

\begin{figure}[t]
    \centering
    \includegraphics[width=\linewidth]{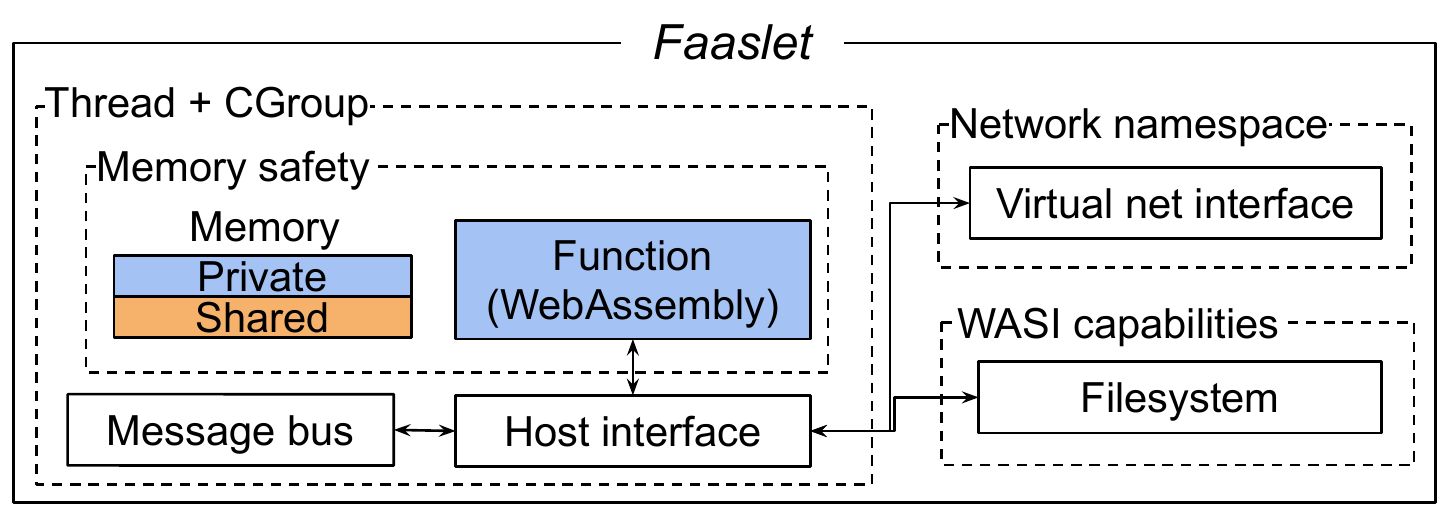}
    \caption{Faaslet abstraction with isolation}
    \label{fig:Faaslet}
\end{figure}

% Host interface table
\begin{table*}[t]    
    \centering\footnotesize
        \begin{tabular}{llll}
        \toprule
            \textbf{Class} & \textbf{Function} & \textbf{Action} & \textbf{Standard} \\    
        \midrule                  
            \multirow{5}{*}{Calls} & \code{byte* \textbf{read\_call\_input}()} & Read input data to function as byte array & \multirow{14}{*}{\emph{none}}\\
            & \code{void \textbf{write\_call\_output}(out\_data)} & Write output data for function & \\
            & \code{int \textbf{chain\_call}(name, args)} & Call function and return the \code{call\_id} & \\
            & \code{int \textbf{await\_call}(call\_id)} & Await the completion of \code{call\_id} & \\
            & \code{byte* \textbf{get\_call\_output}(call\_id)} & Load the output data of \code{call\_id} & \\
        \cmidrule(lr){1-3}
            \multirow{9}{*}{State} & \code{byte* \textbf{get\_state}(key, flags)} & Get pointer to state value for \code{key} & \\
            & \code{byte* \textbf{get\_state\_offset}(key, off, flags)} & Get pointer to state value for \code{key} at offset & \\
            & \code{void \textbf{set\_state}(key, val)} & Set state value for \code{key} & \\
            & \code{void \textbf{set\_state\_offset}(key, val, len, off)} & Set \code{len} bytes of state value at offset for \code{key} & \\
            & \code{void \textbf{push/pull\_state}(key)} & Push/pull global state value for \code{key} & \\
            & \code{void \textbf{push/pull\_state\_offset}(key, off)} & Push/pull global state value for \code{key} at offset & \\
            & \code{void \textbf{append\_state}(key, val)} & Append data to state value for \code{key} & \\
            & \code{void \textbf{lock\_state\_read/write}(key)} & Lock local copy of state value for \code{key} & \\
            & \code{void \textbf{lock\_state\_global\_read/write}(key)} & Lock state value for \code{key} globally & \\        
        \midrule
            \multirow{2}{*}{Dynlink} & \code{void* \textbf{dlopen}/\textbf{dlsym}(}...\code{)} & Dynamic linking of libraries & \multirow{5}{*}{POSIX}\\
            & \code{int \textbf{dlclose}(}...\code{)} & \emph{As above} & \\
        \cmidrule(lr){1-3}
            \multirow{2}{*}{Memory} & \code{void* \textbf{mmap}(}...\code{)}, \code{int \textbf{munmap}(}...\code{)} & Memory grow/shrink only & \\
            & \code{int \textbf{brk}(}...\code{)}, \code{void* \textbf{sbrk}(}...\code{)} & Memory grow/shrink & \\        
        \midrule
            \multirow{2}{*}{Network} & \code{int \textbf{socket}/\textbf{connect}/\textbf{bind}(}...\code{)} & Client-side networking only & \multirow[c]{7}{*}{WASI} \\    
            & \code{size\_t \textbf{send}/\textbf{recv}(}...\code{)} & Send/recv via virtual interface & \\    
        \cmidrule(lr){1-3}
            \multirow{2}{*}{File I/O} & \code{int \textbf{open}/\textbf{close}/\textbf{dup}/\textbf{stat}(}...\code{)} & Per-user virtual filesystem access & \\
            & \code{size\_t \textbf{read}/\textbf{write}(}...\code{)} & \emph{As above} & \\        
        \cmidrule(lr){1-3}
            \multirow{2}{*}{Misc} & \code{int \textbf{gettime}(}...\code{)} & Per-user monotonic clock only & \\      
            & \code{size\_t \textbf{getrandom}(}...\code{)} & Uses underlying host \code{/dev/urandom} & \\     
        \bottomrule
        \end{tabular}
        \caption{Faaslet host interface \textnormal{(The final column indicates
            whether functions are defined as part of POSIX or 
            WASI~\cite{WASIWebsite}.)}}
        \label{table:HostInterface}   
\end{table*}

% Functional requirements
We propose \emph{Faaslets}, a new isolation mechanism that satisfies all the
requirements for efficient data-intensive serverless computing.
\T\ref{table:Alternatives} highlights Faaslets'
strong memory and resource isolation guarantees, and efficient \emph{shared
in-memory state}. Faaslets provide a minimal level of lightweight virtualisation
through their \emph{host interface}, which supports serverless-specific
tasks, memory management, a limited filesystem and network access.

In terms of non-functional requirements, Faaslets improve on containers and VMs
by having a memory footprint below 200\unit{KB} and cold-start initialisation 
times of less than 10\unit{ms}. Faaslets
execute functions compiled to secure IR, allowing them to support multiple
programming languages.

% Proto-Faaslets and initialisation times
While Faaslets cannot initialise as quickly as pure SFI, they mitigate the
cold-start problem through ahead-of-time initialisation from snapshots called 
\emph{Proto-Faaslets}. Proto-Faaslets reduce initialisation times to hundreds of
microseconds, and a single snapshot can be restored across hosts, quickly 
scaling horizontally on clusters.

\subsection{Overview}
\label{sec:faaslet:overview}

% Memory isolation and wasm
\F\ref{fig:Faaslet} shows a function isolated inside a Faaslet. The function
itself is compiled to WebAssembly~\cite{wasm2017}, guaranteeing memory safety
and control flow integrity. By default, a function is placed in its own
\emph{private} contiguous memory region, but Faaslets also support \emph{shared
regions} of memory (\S\ref{sec:faaslet:shared_memory}). This allows a Faaslet to
access shared in-memory state within the constraints of WebAssembly's memory
safety guarantees.

% Resource isolation - CPU
Faaslets also ensure fair resource access. For CPU isolation, they use the CPU
subset of Linux \emph{cgroups}~\cite{LinuxKernelWebsite}. Each function is
executed by a dedicated \emph{thread} of a shared runtime process. This thread
is assigned to a cgroup with a share of CPU equal to that of all Faaslets. The
Linux CFS~\cite{LinuxKernelWebsite} ensures that these threads are
scheduled with equal CPU time.

% Resource isolation - network
Faaslets achieve secure and fair network access using \emph{network
  namespaces}, \emph{virtual network interfaces} and \emph{traffic
  shaping}~\cite{LinuxKernelWebsite}. Each Faaslet has its own network
interface in a separate namespace, configured using \emph{iptables}
rules. To ensure fairness between co-located tenants, each Faaslet applies
traffic shaping on its virtual network interface using \code{tc}, thus
enforcing ingress and egress traffic rate limits.

% Host interface 
As functions in a Faaslet must be permitted to invoke standard system calls to
perform memory management and I/O operations, Faaslets offer an
interface through which to interact with the underlying host. Unlike containers
or VMs, Faaslets do not provide a fully-virtualised POSIX environment but
instead support a minimal serverless-specific host interface
(see~\F\ref{fig:Faaslet}). Faaslets virtualise system calls that interact with 
the underlying host and expose a range of functionality, as described below.

% Message bus
The host interface integrates with the serverless runtime through a
\emph{message bus} (see~\F\ref{fig:Faaslet}). The message bus is used by
Faaslets to communicate with their parent process and each other, receive
function calls, share work, invoke and await other functions, and to be told by
their parent process when to spawn and terminate.

% Filesystem
Faaslets support a \emph{read-global write-local} filesystem, which lets
functions read files from a global object store~(\S\ref{sec:arch}),
and write to locally cached versions of the files. This is primarily used
to support legacy applications, notably language runtimes such as
CPython~\cite{CPythonGithub}, which need a filesystem for loading library code
and storing intermediate bytecode. The filesystem is accessible through a set of
POSIX-like API functions that implement the WASI capability-based security
model, which provides efficient isolation through unforgeable file
handles~\cite{WASIDesignWebsite}. This removes the need for more
resource-intensive filesystem isolation such as a layered filesystem or
\code{chroot}, which otherwise add to cold
start latencies~\cite{SOCK:2018}.

\subsection{Host Interface}
\label{sec:faaslet:host_interface}

% Host interface requirements and risks
The Faaslet host interface must provide a virtualisation layer capable
of executing a range of serverless big data applications, as well as legacy
POSIX applications. This interface
necessarily operates outside the bounds of memory safety, and hence is trusted
to preserve isolation when interacting with the host.

% Critique of existing approaches
In existing serverless platforms based on containers and VMs, this
virtualisation layer is a standard POSIX environment, with serverless-specific
tasks executed through language- and provider-specific APIs over
HTTP~\cite{AWSLambdaWebsite, GCFWebsite, IBMCloudFunctionsWebsite}.
Instantiating a full POSIX environment with the associated isolation
mechanisms leads to high initialisation times~\cite{SOCK:2018}, and heavy use of HTTP APIs contributes further latency and
network overheads.

% Summary of Faaslet host interface
In contrast, the Faaslet host interface targets minimal virtualisation, hence
reducing the overheads required to provide isolation. The host interface is a
low-level API built exclusively to support a range of high-performance
serverless applications. The host interface is dynamically linked with
function code at runtime (\S\ref{sec:faaslet:building}), making calls to the
interface more efficient than performing the same tasks through an external 
API. 

% Table
\T\ref{table:HostInterface} lists the Faaslet host interface API, which
supports: (i) chained serverless function invocation; (ii) interacting with
shared state {(\S\ref{sec:state})}; (iii) a subset of POSIX-like calls for
memory management, timing, random numbers, file/network I/O and dynamic linking.
A subset of these POSIX-like calls are implemented according to WASI, an
emerging standard for a server-side WebAssembly interface~\cite{WASIWebsite}.
Some key details of the API are as follows:

% Chaining functions
\mypar{Function invocation} Functions retrieve their input data serialised as
byte arrays using the \code{read\_call\_input} function, and similarly write
their output data as byte arrays using \code{write\_call\_output}. Byte arrays
constitute a generic, language-agnostic interface.

Non-trivial serverless applications invoke multiple functions that work
together as part of chained calls, made with the \code{chain\_call}
function. Users' functions have unique names, which are passed to
\code{chain\_call}, along with a byte array containing the input data for that
call.

A call to \code{chain\_call} returns the call ID of the invoked function. The
call ID can then be passed to \code{await\_call} to perform a blocking wait for
another call to finish or fail, yielding its return code. The Faaslet blocks
until the function has completed, and passes the same call ID to
\code{get\_call\_output} to retrieve the chained call's output data.

Calls to \code{chain\_call} and \code{await\_call} can be used in loops to
spawn and await calls in a similar manner to standard multi-threaded code: one
loop invokes \code{chain\_call} and records the call IDs; a second loop calls
\code{await\_call} on each ID in turn. We show this pattern in Python in
Listing~\ref{lst:sgd}.

% Dynamic linking 
\noindent \mypar{Dynamic linking} Some legacy applications and
libraries require support for dynamic linking, \eg CPython dynamically
links Python extensions. All dynamically loaded code must first be compiled to
WebAssembly and undergo the same validation process as other user-defined
code~(\S\ref{sec:faaslet:building}). Such modules are loaded via the standard
Faaslet filesystem abstraction and covered by the same safety guarantees as its
parent function. Faaslets support this through a standard POSIX dynamic linking
API, which is implemented according to WebAssembly dynamic linking
conventions~\cite{WasmDynamicLinking}.

% Memory
\mypar{Memory} Functions allocate memory dynamically through calls to
\code{mmap()} and \code{brk()}, either directly or through
\code{dlmalloc}~\cite{dlmallocWebsite}. The Faaslet allocates memory in its
private memory region, and uses \code{mmap} on the underlying host to extend
the region if necessary. Each function has its own predefined memory limit, and
these calls fail if growth of the private region would exceeded this limit.

% Networking
\mypar{Networking} The supported subset of networking calls allows simple
client-side send/receive operations and is sufficient for common use cases,
such as connecting to an external data store or a remote HTTP endpoint.  The
functions \code{socket}, \code{connect} and \code{bind} allow setting up the
socket while \code{read} and \code{write} allow the sending and receiving of
data. Calls fail if they pass flags that are not related to simple send/receive
operations over IPv4/IPv6, \eg the \code{AF\_UNIX} flag.

The host interface translates these calls to equivalent socket operations on
the host. All calls interact exclusively with the Faaslet's virtual network
interface, thus are constrained to a private network interface and
cannot exceed rate limits due to the traffic shaping rules.

% Byte arrays
\noindent \mypar{Byte arrays} Function inputs, results and state are
represented as simple byte arrays, as is all function memory. This avoids the
need to serialise and copy data as it passes through the API, and makes it
trivial to share arbitrarily complex in-memory data structures.

\subsection{Shared Memory Regions}
\label{sec:faaslet:shared_memory}

% Summary and benefits
As discussed in \S\ref{sec:background}, sharing in-memory state while otherwise
maintaining isolation is an important requirement for efficient serverless big
data applications. Faaslets do this by adding the new concept of
\emph{shared regions} to the existing WebAssembly memory model~\cite{wasm2017}. 
Shared regions give functions concurrent
access to disjoint segments of shared process memory, allowing them direct,
low-latency access to shared data structures. Shared regions are backed by
standard OS virtual memory, so there is no extra serialisation or overhead,
hence Faaslets achieve efficient concurrent access on a par with native
multi-threaded applications. In \S\ref{sec:state:tiers}, we describe how
Faaslets use this mechanism to provide shared in-memory access to global state.

% Implementation summary
Shared regions maintain the memory safety guarantees of the existing WebAssembly
memory model, and use standard OS virtual memory mechanisms. WebAssembly
restricts each function's memory to a contiguous linear byte array, which is
allocated by the Faaslet at runtime from a disjoint section of the process
memory. To create a new shared region, the Faaslet extends the function's linear
byte array, and remaps the new pages onto a designated region of common process
memory. The function accesses the new region of linear memory as normal, hence
maintaining memory safety, but the underlying memory accesses are mapped onto
the shared region.

\begin{figure}[tb]
    \centering
    \includegraphics[width=0.7\linewidth]{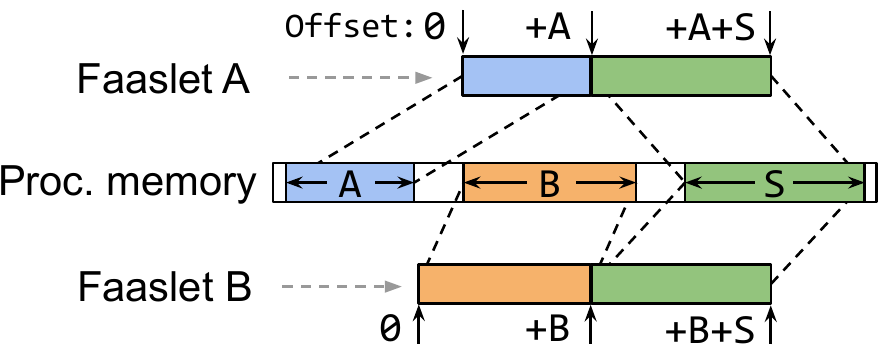}
    \caption{Faaslet shared memory region mapping}
    \label{fig:LocalState}
\end{figure}

% Figure
\F\ref{fig:LocalState} shows Faaslets A and B accessing a shared region
(labelled S), allocated from a disjoint region of the common process memory
(represented by the central region). Each Faaslet has its own region of private
memory~(labelled A and B), also allocated from the process memory. Functions
inside each Faaslet access all memory as offsets from zero, forming a single
linear address space. Faaslets map these offsets onto either a private region
(in this case the lower offsets), or a shared region (in this case the higher
offsets). 

Multiple shared regions are permitted, and functions can also extend their
private memory through calls to the memory management functions in the host
interface such as \code{brk}~(\S\ref{sec:faaslet:host_interface}). Extension of
private memory and creation of new shared regions is handled by extending a
byte array, which represents the function's memory, and then remapping the underlying
pages to regions of shared process memory. This means the function
continues to see a single densely-packed linear address space, which may be
backed by several virtual memory mappings. Faaslets allocate 
shared process memory through calls to \code{mmap} on the underlying host, passing \code{MAP\_SHARED}
and \code{MAP\_ANONYMOUS} flags to create shared and private regions,
respectively, and remap these regions with \code{mremap}.

\subsection{Building Functions for Faaslets} \label{sec:faaslet:building}

\begin{figure}[t]
    \centering
    \includegraphics[width=1.0\linewidth]{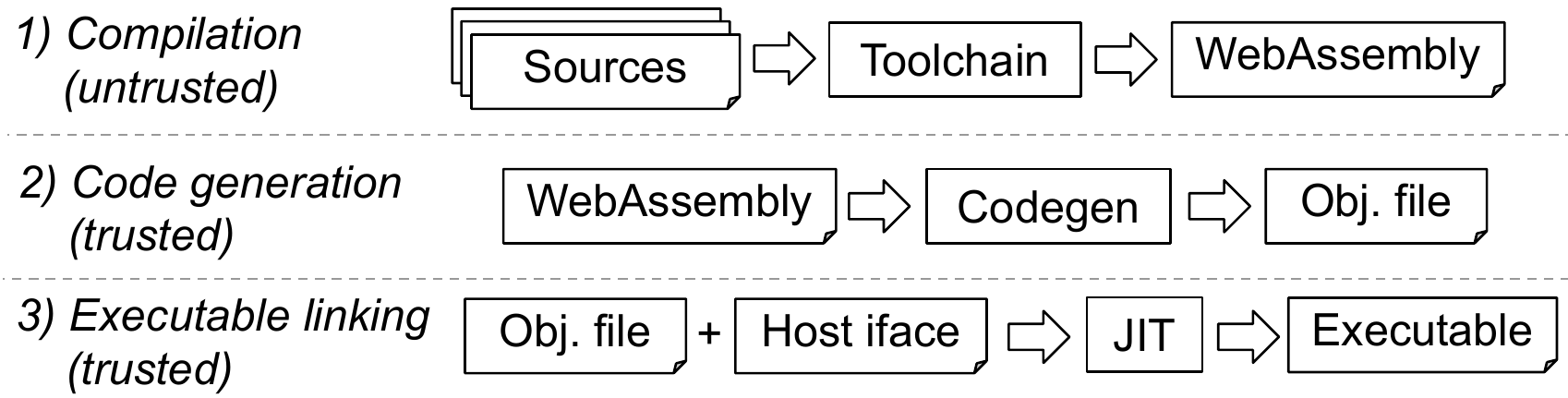}
    \caption{Creation of a Faaslet executable}
    \label{fig:FaasmCodegen}
\end{figure}

\F\ref{fig:FaasmCodegen} shows the three phases to convert source code of a
function into a Faaslet executable: (1)~the user invokes the Faaslet toolchain
to compile the function into a WebAssembly binary, linking against a
language-specific declaration of the Faaslet host interface; (2)~code
generation creates an object file with machine code from WebAssembly; and
(3)~the host interface definition is linked with the machine code to produce
the Faaslet executable.

When Faaslets are deployed, the compilation phase to generate the WebAssembly
binary takes place on a user's machine. Since that is untrusted, the code
generation phase begins by validating the WebAssembly binary, as defined in the
WebAssembly specification~\cite{wasm2017}. This ensures that the binary
conforms to the specification. Code generation then takes place in a trusted
environment, after the user has uploaded their function.

In the linking phase, the Faaslet uses LLVM JIT
libraries~\cite{LLVM9Website} to link the object file and the definition of
the host interface implementation. The host interface functions are defined as
\emph{thunks}, which allows injecting the trusted host interface implementation
into the function binary.

Faaslets use WAVM~\cite{WAVMGithub} to perform the validation, code generation
and linking. WAVM is an open-source WebAssembly VM, which passes the
WebAssembly conformance tests~\cite{MechanisingWasm:2018} and thus guarantees
that the resulting executable enforces memory safety and control flow
integrity~\cite{wasm2017}.

% End

%% file: sections/state.tex
% State

\section{Local and Global State} \label{sec:state}

\noindent Stateful serverless applications can be created with Faaslets using
\emph{distributed data objects (DDO)}, which are language-specific classes that
expose a convenient high-level state interface. DDOs are implemented using the 
key/value state API from \T\ref{table:HostInterface}.

The state associated with Faaslets is managed using a \emph{two-tier} approach
that combines local sharing with global distribution of state: a \emph{local
  tier} provides shared in-memory access to state on the same host; and a
\emph{global tier} allows Faaslets to synchronise state across hosts.

DDOs hide the two-tier state architecture, providing transparent access to
distributed data. Functions, however, can still access the state API directly,
either to exercise more fine-grained control over consistency and
synchronisation, or to implement custom data structures.

\begin{lstlisting}[float=t,label={lst:sgd},language=Python, caption=Distributed
  SGD application with Faaslets]
t_a = (*@\textbf{SparseMatrixReadOnly("training\_a")}@*) (*@\label{lst:sgd:training1}@*)
t_b = (*@\textbf{MatrixReadOnly("training\_b")}@*) (*@\label{lst:sgd:training2}@*)
weights = (*@\textbf{VectorAsync("weights")}@*) (*@\label{lst:sgd:weights}@*)

@faasm_func(*@\label{lst:sgd:annotation1}@*)
def weight_update(idx_a, idx_b): (*@\label{lst:sgd:weight_update}@*)
  for col_idx, col_a in (*@\textbf{t\_a.columns[idx\_a:idx\_b]}@*): (*@\label{lst:sgd:input1}@*)
    col_b = (*@\textbf{t\_b.columns[col\_idx]}@*) (*@\label{lst:sgd:input2}@*)
    adj = calc_adjustment(col_a, col_b)
    for val_idx, val in col_a.non_nulls():        
      (*@\textbf{weights[val\_idx] += val * adj}@*) (*@\label{lst:sgd:update}@*)
      if iter_count % threshold == 0:
        (*@\textbf{weights.push()}@*) (*@\label{lst:sgd:push}@*)
                
@faasm_func(*@\label{lst:sgd:annotation2}@*)
def sgd_main(n_workers, n_epochs): (*@\label{lst:sgd:sgd_main}@*)
  for e in n_epochs:
    args = divide_problem(n_workers)
    c = (*@\textbf{chain(update, n\_workers, args)}@*) (*@\label{lst:sgd:chain}@*)
    (*@\textbf{await\_all(c)}@*)
    (*@\textnormal{...}@*)
\end{lstlisting}

\subsection{State Programming Model}

Each DDO represents a single state value, referenced throughout the system 
using a string holding its respective state key.

% Push and pull
Faaslets write changes from the local to the global tier by performing a
\emph{push}, and read from the global to the local tier by performing a
\emph{pull}. DDOs may employ push and pull operations to produce variable
consistency, such as delaying updates in an eventually-consistent list or set,
and may lazily pull values only when they are accessed, such as in a distributed
dictionary. Certain DDOs are immutable, and hence avoid repeated
synchronisation.

% Listing 
Listing~\ref{lst:sgd} shows both implicit and explicit use of two-tier
state through DDOs to implement stochastic gradient descent~(SGD) in Python.
The
\code{weight\_update} function accesses two large input matrices through the
\code{SparseMatrixReadOnly} and
\code{MatrixReadOnly} DDOs (lines~\ref{lst:sgd:training1}
and~\ref{lst:sgd:training2}), and a single shared weights vector using
\code{VectorAsync} (line~\ref{lst:sgd:weights}). \code{VectorAsync}
exposes a \code{push()} function which is used to periodically push updates from
the local tier to the global tier (line~\ref{lst:sgd:push}). The calls to 
\code{weight\_update} are chained in a loop in \code{sgd\_main} 
(line~\ref{lst:sgd:chain}).

Function \code{weight\_update} accesses a randomly assigned subset of columns
from the training matrices using the \code{columns} property
(lines~\ref{lst:sgd:input1} and~\ref{lst:sgd:input2}). The DDO implicitly
performs a pull operation to ensure that data is present, and only replicates 
the necessary subsets of the state values in the
local tier---the entire matrix is not transferred unnecessarily.

Updates to the shared weights vector in the local tier are made in a loop in
the \code{weight\_update} function (line~\ref{lst:sgd:update}). It invokes the
\code{push} method on this vector (line~\ref{lst:sgd:push}) sporadically to
update the global tier. This improves performance and reduces network overhead,
but introduces inconsistency between the tiers. SGD tolerates such
inconsistencies and it does not affect the overall result.

\subsection{Two-Tier State Architecture} \label{sec:state:tiers}

\begin{figure}[t]
  \centering \includegraphics[width=\linewidth]{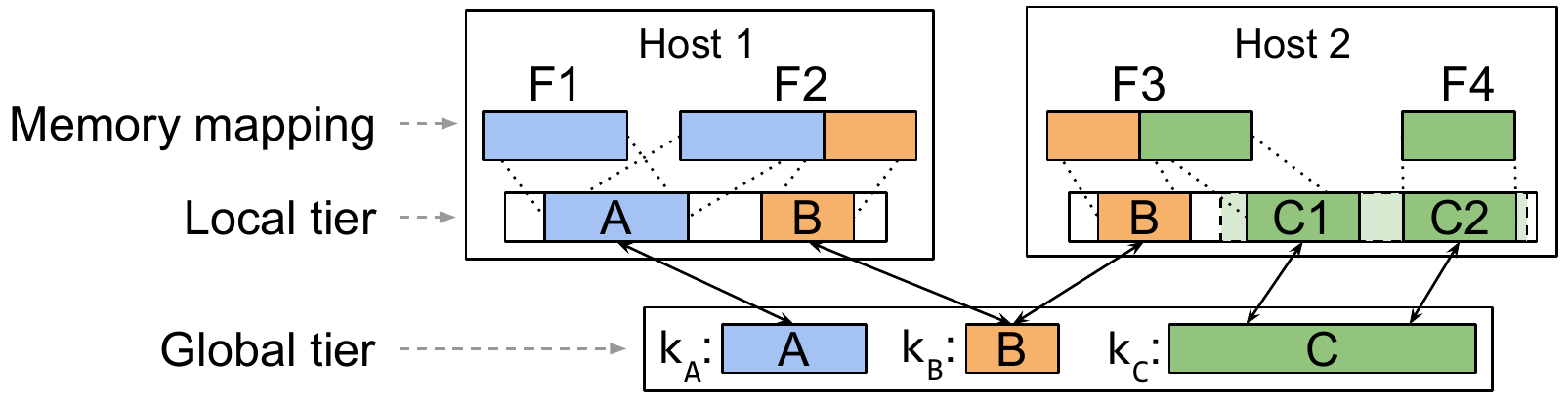}
  \caption{Faaslet two-tier state architecture}
  \label{fig:RemoteState}
\end{figure}

% Summary
Faaslets represent state with a key/value abstraction, using unique \emph{state
 keys} to reference \emph{state values}. The authoritative state value for each
 key is held in the global tier, which is backed by a distributed
 key-value store (KVS) and accessible to all Faaslets in the cluster.
 Faaslets on a given host share a local
 tier, containing replicas of each state value currently mapped to Faaslets on
 that host. The local tier is held exclusively in Faaslet shared memory
 regions, and Faaslets do not have a separate local storage service, as in
 SAND~\cite{SAND:2018} or Cloudburst~\cite{Cloudburst:2020}.
 
\F\ref{fig:RemoteState} shows the two-tier state architecture across two
hosts. Faaslets on host~1 share state value~A; Faaslets on both hosts share
state value~B. Accordingly, there is a replica of state value~A in the local
tier of host~1, and replicas of state value~B in the local tier of both hosts.

The \code{columns} method of the \code{SparseMatrixReadOnly} and
\code{MatrixReadOnly} DDOs in Listing~\ref{lst:sgd} uses
\emph{state chunks} to access a subset of a larger state value. As shown in
\F\ref{fig:RemoteState}, state value~C has state chunks, which are treated as
smaller independent state values. Faaslets create replicas of only the required
chunks in their local tier.

\mypar{Ensuring local consistency} State value replicas in the local tier are
created using Faaslet shared memory (\S\ref{sec:faaslet:shared_memory}). To
ensure consistency between Faaslets accessing a replica, Faaslets acquire a
\emph{local read lock} when reading, and a \emph{local write lock} when
writing. This locking happens implicitly as part of all state API functions,
but not when functions write directly to the local replica via a pointer. The
state API exposes the \code{lock\_state\_read} and \code{lock\_state\_write}
functions that can be used to acquire local locks explicitly, \eg to implement
a list that performs multiple writes to its state value when atomically adding
an element. A Faaslet creates a new local replica after a call to
\code{pull\_state} or \code{get\_state} if it does not already exist, and
ensures consistency through a write lock.

\mypar{Ensuring global consistency} DDOs can produce varying levels of
consistency between the tiers as shown by \code{VectorAsync} in
Listing~\ref{lst:sgd}. To enforce strong consistency, DDOs must use \emph{global
read/write locks}, which can be acquired and released for each state key using
\code{lock\_state\_global\_read} and \code{lock\_state\_global\_write}, respectively. To perform a consistent write to the global tier, an
object acquires a global write lock, calls \code{pull\_state} to update the
local tier, applies its write to the local tier, calls \code{push\_state} to
update the global tier, and releases the lock.

% End

%% file: sections/arch.tex
% Faasm

\section{F{\small{}AASM} Runtime}
\label{sec:arch}

\begin{figure}[t]
\centering
\includegraphics[width=0.6\linewidth]{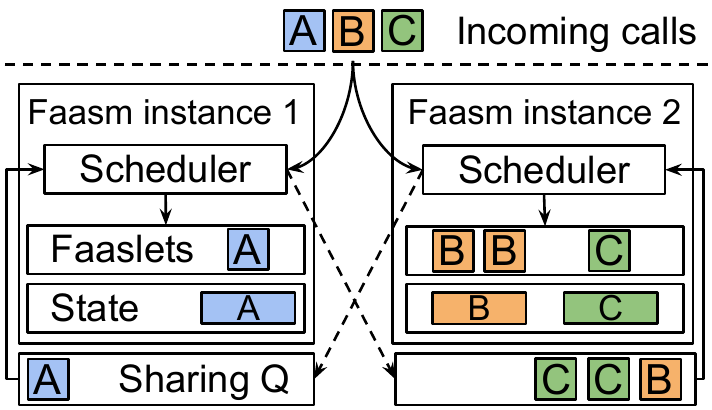}
\caption{\sys system architecture}
\label{fig:FaasmArch}
\end{figure}

\sys is the serverless runtime that uses Faaslets to execute distributed
stateful serverless applications across a cluster. \sys is designed to
integrate with existing serverless platforms, which provide the underlying
infrastructure, auto-scaling functionality and user-facing frontends. \sys
handles the scheduling, execution and state management of Faaslets. The design
of \sys follows a distributed architecture: multiple \sys runtime instances
execute on a set of servers, and each instance manages a pool of Faaslets.

\subsection{Distributed Scheduling}
\label{sec:arch:scheduling}

A local scheduler in the \sys runtime is responsible for the scheduling of
Faaslets. Its scheduling strategy is key to minimising
data-shipping (see~\S\ref{sec:background}) by ensuring that executed functions
are co-located with required in-memory state. One or more Faaslets managed by a
runtime instance may be \emph{warm}, \ie they already have their code and state
loaded. The scheduling goal is to ensure that as many function calls as possible
are executed by warm Faaslets.

To achieve this without modifications to the underlying platform's scheduler,
\sys uses a \emph{distributed shared state} scheduler similar to
Omega~\cite{Omega:2013}. Function calls are sent round-robin to local
schedulers, which execute the function locally if they are warm and have
capacity, or share it with another warm host if one exists. The set of warm
hosts for each function is held in the \sys state global tier, and each
scheduler may query and atomically update this set during the scheduling
decision.

\F\ref{fig:FaasmArch} shows two \sys runtime instances, each with its own local
scheduler, a pool of Faaslets, a collection of state stored in memory, and a
sharing queue. Calls for functions~A--C are received by the local schedulers,
which execute them locally if they have warm Faaslets, and share them with the
other host if not. Instance~1 has a warm Faaslet for function~A and accepts
calls to this function, while sharing calls to functions~B and C with
Instance~2, which has corresponding warm Faaslets. If a function call is
received and there are no instances with warm Faaslets, the instance that
received the call creates a new Faaslet, incurring a ``cold start''.

\subsection{Reducing Cold Start Latency}
\label{sec:arch:proto}

While Faaslets typically initialise in under 10\unit{ms}, \sys reduces this
further using \emph{Proto-Faaslets}, which are Faaslets that contain snapshots
of arbitrary execution state that can be restored on any host in the cluster.
From this snapshot, \sys spawns a new Faaslet instance, typically
reducing initialisation to hundreds of microseconds
(\S\ref{sec:eval:overheads}).

% User-defined initialisation code
Different Proto-Faaslets are generated for a function by specifying
user-defined \emph{initialisation code}, which is executed before snapshotting.
If a function executes the same code on each invocation, that code can become
initialisation code and be removed from the function itself.
For Faaslets with dynamic language runtimes, the runtime initialisation can be
done as part of the initialisation code.

% Snapshot details
A Proto-Faaslet snapshot includes a function's stack, heap, function table,
stack pointer and data, as defined in the WebAssembly
specification~\cite{wasm2017}. Since WebAssembly memory is represented by a
contiguous byte array, containing the stack,heap and data, \sys restores a
snapshot into a new Faaslet using a copy-on-write memory mapping.  All other
data is held in standard C++ objects.  Since the snapshot is independent of the
underlying OS thread or process, \sys can serialise Proto-Faaslets and
instantiate them across hosts.

% Uploading
\sys provides an \emph{upload} service that exposes an HTTP endpoint. Users
upload WebAssembly binaries to this endpoint, which then performs code
generation~(\S\ref{sec:faaslet:building}) and writes the resulting object files
to a shared \emph{object store}. The implementation of this store is specific to
the underlying serverless platform but can be a cloud provider's own solution
such as AWS S3~\cite{AWSS3Website}. Proto-Faaslets are generated and stored in
the \sys global state tier as part of this process. When a Faaslet undergoes a
cold start, it loads the object file and Proto-Faaslet, and restores it.

In addition, \sys uses Proto-Faaslets to reset Faaslets after each function
call. Since the Proto-Faaslet captures a function's initialised execution
state, restoring it guarantees that no information from the previous call is
disclosed. This can be used for functions that are \emph{multi-tenant}, \eg in
a serverless web application. \sys guarantees that private data held in memory
is cleared away after each function execution, thereby allowing Faaslets to
handle subsequent calls across tenants. In a container-based platform, this is
typically not safe, as the platform cannot ensure that the container memory has
been cleaned entirely between calls.

% End

%% file: sections/evaluation.tex
% Evaluation

\section{Evaluation} \label{sec:eval}
  
Our experimental evaluation targets the following questions: (i)~how does \sys
state management improve efficiency and performance on parallel machine learning
training?~(\S\ref{sec:eval:training}) (ii)~how do Proto-Faaslets and low
initialisation times impact performance and throughput in inference
serving?~(\S\ref{sec:eval:inference}) (iii)~how does Faaslet isolation affect
performance in a linear algebra benchmark using a dynamic language
runtime?~(\S\ref{sec:eval:python}) and
(iv)~how do the overheads of Faaslets compare to Docker
containers?~(\S\ref{sec:eval:overheads})

\subsection{Experimental Set-up}

\mypar{Serverless baseline} To benchmark \sys against a state-of-the-art
serverless platform, we use Knative~\cite{KNativeGithub}, a container-based
system built on Kubernetes~\cite{KubernetesWebsite}. All experiments are
implemented using the same code for both \sys and Knative, with a
Knative-specific implementation of the Faaslet host interface for
container-based code. This interface uses the same undelrying state management
code as \sys, but cannot share the local tier between co-located functions.
Knative function chaining is performed through the standard Knative API. Redis 
is used for the distributed KVS and deployed to the same cluster.

\mypar{\sys integration} We integrate \sys with Knative by running \sys runtime
instances as Knative functions that are replicated using the default
autoscaler. The system is otherwise unmodified, using the default endpoints and
scheduler.

\mypar{Testbed} Both \sys and Knative applications are executed on the same
Kubernetes cluster, running on 20~hosts, all Intel Xeon E3-1220
3.1\unit{GHz} machines with 16\unit{GB} of RAM, connected with a 1\unit{Gbps}
connection. Experiments in \S\ref{sec:eval:overheads} were run on a single 
Intel Xeon E5-2660 2.6\unit{GHz} machine with 32\unit{GB} of
RAM.

\mypar{Metrics} In addition to the usual evaluation metrics, such as execution
time, throughput and latency, we also consider \emph{billable memory}, which
quantifies memory consumption over time. It is the product of the peak function
memory multiplied by the number and runtime of functions, in units of
GB-seconds. It is used to attribute memory usage in many serverless
platforms~\cite{AWSLambdaWebsite, GCFWebsite, IBMCloudFunctionsWebsite}. Note
that all memory measurements include the containers/Faaslets and their state.

\subsection{Machine Learning Training} \label{sec:eval:training}

\begin{figure*}[t]    
  \begin{subfigure}[]{0.32\textwidth}
      \includegraphics[width=\textwidth]{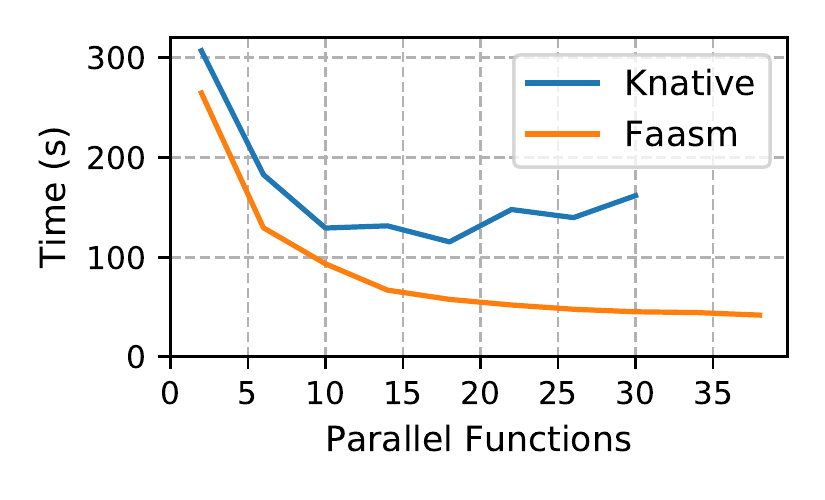}            
      \caption{Training time}
      \label{fig:SGDDuration}
  \end{subfigure}%
  ~
  \begin{subfigure}[]{0.32\textwidth}
      \includegraphics[width=\textwidth]{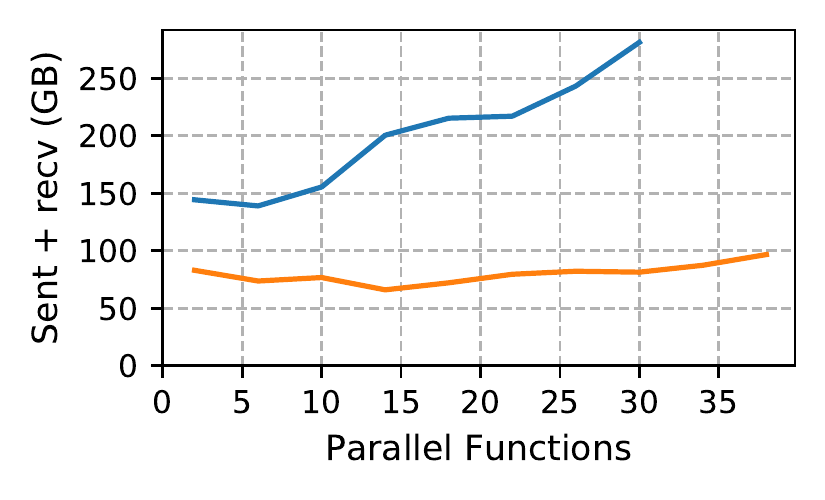}            
      \caption{Network transfers}
      \label{fig:SGDNetwork}
  \end{subfigure}%
  ~
  \begin{subfigure}[]{0.32\textwidth}
    \includegraphics[width=\textwidth]{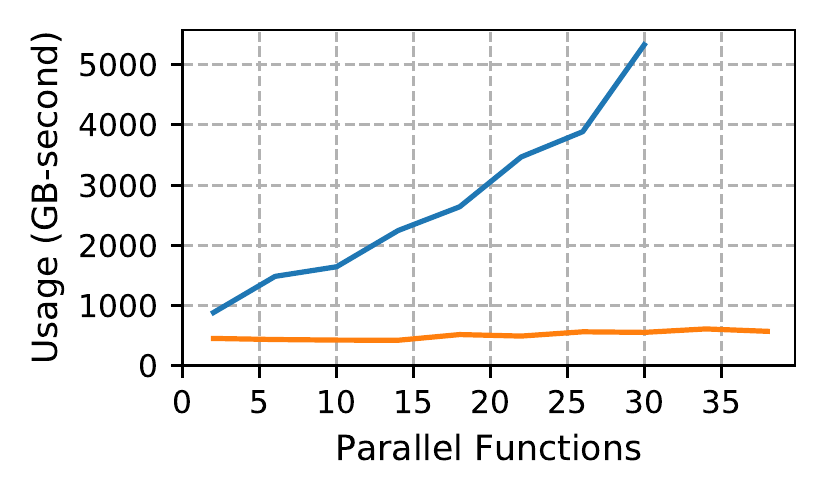}            
    \caption{Memory usage}
    \label{fig:SGDMemory}
  \end{subfigure}
  \caption{Machine learning training with SGD with Faaslets~(\sys) and
    containers~(Knative)}
  \label{fig:SGDResults}
\end{figure*}

% Goal
This experiment focuses on the impact of \sys{}'s state management on runtime,
network overheads and memory usage.

% Set-up
We use distributed \emph{stochastic gradient descent}~(SGD) using the
\textsc{Hogwild!} algorithm~\cite{Hogwild:2011} to run text classification on
the Reuters RCV1 dataset~\cite{lewis2004rcv1}. This updates a central weights
vector in parallel with batches of functions across multiple epochs. We run
both Knative and \sys with increasing numbers of parallel functions.

% Results
\F\ref{fig:SGDDuration} shows the training time. \sys exhibits a small
improvement in runtime of 10\% compared to Knative at low parallelism and a
60\% improvement with 15~parallel functions. With more than 20~parallel Knative
functions, the underlying hosts experience increased memory pressure and they
exhaust memory with over 30~functions. Training time continues to improve for
\sys up to 38~parallel functions, at which point there is a more than an 80\%
improvement over 2~functions.

\F\ref{fig:SGDNetwork} shows that, with increasing parallelism, the volume of
network transfers increases in both \sys and Knative. Knative transfers more
data to start with and the volume increase more rapidly, with 145\unit{GB}
transferred with 2~parallel functions and 280\unit{GB} transferred with
30~functions. \sys transfers 75\unit{GB} with 2~parallel functions and
100\unit{GB} with 38~parallel functions.

\F\ref{fig:SGDMemory} shows that billable memory in Knative increases with more
parallelism: from 1,000\unit{GB-seconds} for 2~functions to over
5,000\unit{GB-second} for 30~functions. The billable memory for \sys increases
slowly from 350\unit{GB-second} for 2~functions to 500\unit{GB-second} with
38~functions.

% Interpretation
The increased network transfer, memory usage and duration in Knative is caused
primarily by data shipping, \eg loading data into containers. \sys benefits
from sharing data through its local tier, hence amortises overheads and reduces
latency. Further improvements in duration and network overhead come from
differences in the updates to the shared weights vector: in \sys, the updates
from multiple functions are batched per host; whereas in Knative, each function
must write directly to external storage. Billable memory in Knative and \sys
increases with more parallelism, however, the increased memory footprint and
duration in Knative make this increase more pronounced.

% Micro dataset
To isolate the underlying performance and resource overheads of \sys and
Knative, we run the same experiment with the number of training examples reduced
from 800K to 128. Across 32 parallel functions, we observe for \sys and
Knative: training times of 460\unit{ms} and 630\unit{ms}; network
transfers of 19\unit{MB} and 48\unit{MB}; billable memory usage of
0.01\unit{GB-second} and 0.04\unit{GB-second}, respectively.

% Micro dataset interpretation
In this case, increased duration in Knative is caused by the latency and
volume of inter-function communication through the Knative HTTP API versus
direct inter-Faaslet communication. \sys incurs reduced network transfers versus
Knative as in the first experiment, but the overhead of these transfers in both
systems are negligible as they are small and amortized across all functions.
Billable memory is increased in Knative due to the memory overhead of each
function container being 8\unit{MB} (versus 270\unit{kB} for each Faaslet). These
improvements are negligible when compared with those derived from reduced data
shipping and duplication of the full dataset.

\subsection{Machine Learning Inference} \label{sec:eval:inference}

% Aim
This experiment explores the impact of the Faaslet initialisation times on
cold-starts and function call throughput.

% Set-up
We consider a machine learning inference application because they are typically
user-facing, thus latency-sensitive, and must serve high volumes of
requests. We perform inference serving with TensorFlow
Lite~\cite{TFLiteWebsite}, with images loaded from a file server and classified
using a pre-trained MobileNet~\cite{MobileNets:2017} model. In our
implementation, requests from each user are sent to different instances of the
underlying serverless function. Therefore, each user sees a cold-start on their
first request. We measure the latency distribution and change in median latency
when increasing throughput and varying the ratio of cold-starts.

\begin{figure}[t]    
  \begin{subfigure}[]{0.5\linewidth}
      \includegraphics[width=\textwidth]{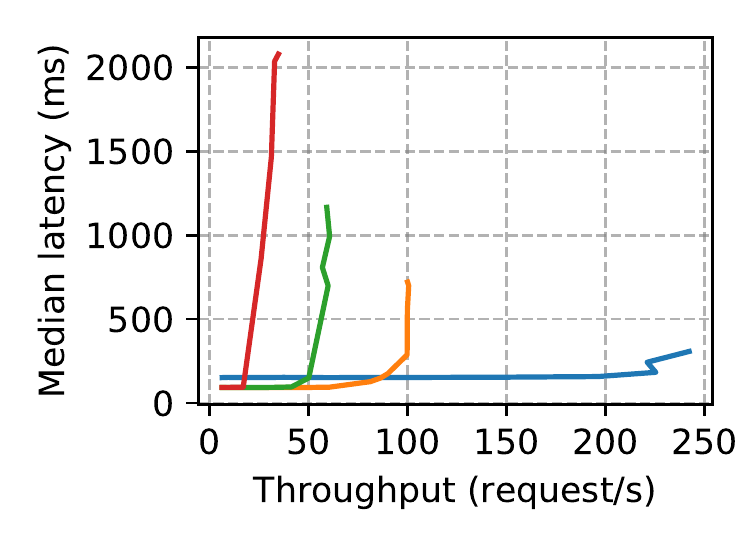}            
      \caption{Throughput vs. latency}
      \label{fig:TFThroughput}
  \end{subfigure}%
  ~
  \begin{subfigure}[]{0.5\linewidth}
      \includegraphics[width=\textwidth]{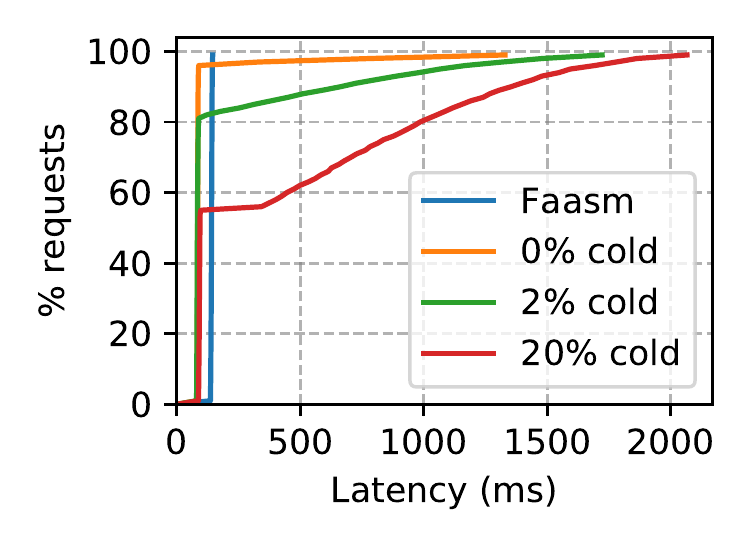}            
      \caption{Latency CDF}
      \label{fig:TFLatency}
  \end{subfigure}
  \caption{Machine learning inference with TensorFlow Lite}
  \label{fig:TFResults}
\end{figure}

\Fs\ref{fig:TFThroughput} and~\ref{fig:TFLatency} show a single line for \sys
that covers all cold-start ratios. Cold-starts only introduce a negligible
latency penalty of less than 1\unit{ms} and do not add significant resource
contention, hence all ratios behave the same. Optimal latency in \sys is higher
than that in Knative, as the inference calculation takes longer due to the
performance overhead from compiling TensorFlow Lite to WebAssembly.

\F\ref{fig:TFThroughput} shows that the median latency in Knative increases
sharply from a certain throughput threshold depending on the cold-start
ratio. This is caused by cold starts resulting in queuing and resource
contention, with the median latency for the 20\% cold-start workload increasing
from 90\unit{ms} to over 2\unit{s} at around 20\unit{req/s}. \sys maintains a
median latency of 120\unit{ms} at a throughput of over 200\unit{req/s}.

\F\ref{fig:TFLatency} shows the latency distribution for a single function that
handles successive calls with different cold-start ratios. Knative has a tail
latency of over 2\unit{s} and more than 35\% of calls have latencies of over
500\unit{ms} with 20\% cold-starts. \sys achieves a tail latency of under
150\unit{ms} for all ratios.

\subsection{Language Runtime Performance with Python} \label{sec:eval:python}

\begin{figure}[t]
  \begin{subfigure}[]{0.5\linewidth}
      \includegraphics[width=\textwidth]{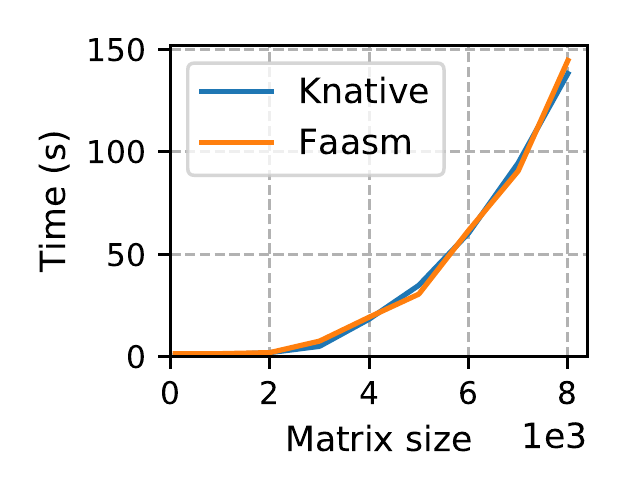}            
      \caption{Duration}
      \label{fig:MatrixDuration}
  \end{subfigure}%
  ~
  \begin{subfigure}[]{0.5\linewidth}
      \includegraphics[width=\textwidth]{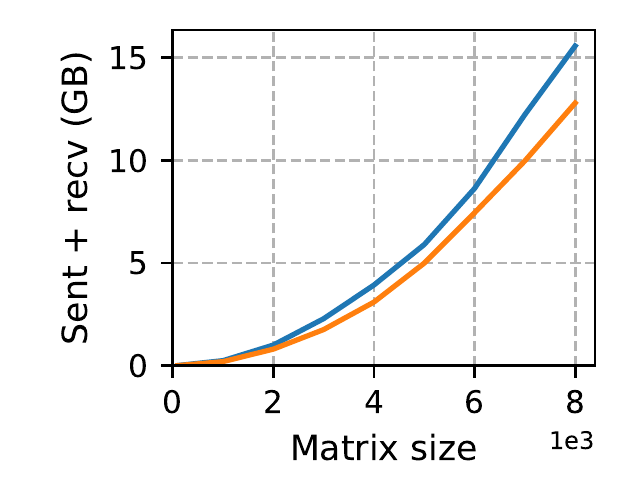}            
      \caption{Network transfer}
      \label{fig:MatrixNetwork}
  \end{subfigure}  
  \caption{Comparison of matrix multiplication with Numpy}
  \label{fig:MatrixResults}
\end{figure}

% Aim
The next two experiments (i)~measure the performance impact of Faaslet isolation
on a distributed benchmark using an existing dynamic language runtime,
the CPython interpreter; and (ii)~investigate
the impact on a single Faaslet running compute microbenchmarks and a suite
of Python microbenchmarks.

% Set up - Python
We consider a distributed divide-and-conquer matrix multiplication implemented
with Python and Numpy. In the \sys implementation, these functions are executed
using CPython inside a Faaslet; in Knative, we use standard Python. As there is
no WebAssembly support for BLAS and LAPACK, we do not use them in either
implementation.

While this experiment is computationally intensive, it also makes use of the
filesystem, dynamic linking, function chaining and state, thus exercising all
of the Faaslet host interface. Each matrix multiplication is subdivided into
multiplications of smaller submatrices and merged. This is implemented by
recursively chaining serverless functions, with each multiplication using
64~multiplication functions and 9~merging functions. We compare the execution
time and network traffic when running multiplications of increasingly large
matrices.

% Results
\F\ref{fig:MatrixDuration} shows that the duration of matrix multiplications on
\sys and Knative are almost identical with increasing matrix sizes. Both take
around 500\unit{ms} with 100$\times$100 matrices, and almost 150\unit{secs}
with 8000$\times$8000 matrices. \F\ref{fig:MatrixNetwork} shows that \sys
results in 13\% less network traffic across all matrix sizes, and hence gains a
small benefit from storing intermediate results more efficiently.

% Set up - microbench
In the next experiment, we use Polybench/C~\cite{PolybenchWebsite} to measure
the Faaslet performance overheads on simple compute functions, and the Python
Performance Benchmarks~\cite{PythonPerformanceWebsite} for overheads on more
complex applications. Polybench/C is compiled directly to WebAssembly and
executed in Faaslets; the Python code executes with CPython running in a
Faaslet.

% Microbenchmark results
\begin{figure}[t]    
  \begin{subfigure}[]{\linewidth}
      \includegraphics[width=\linewidth]{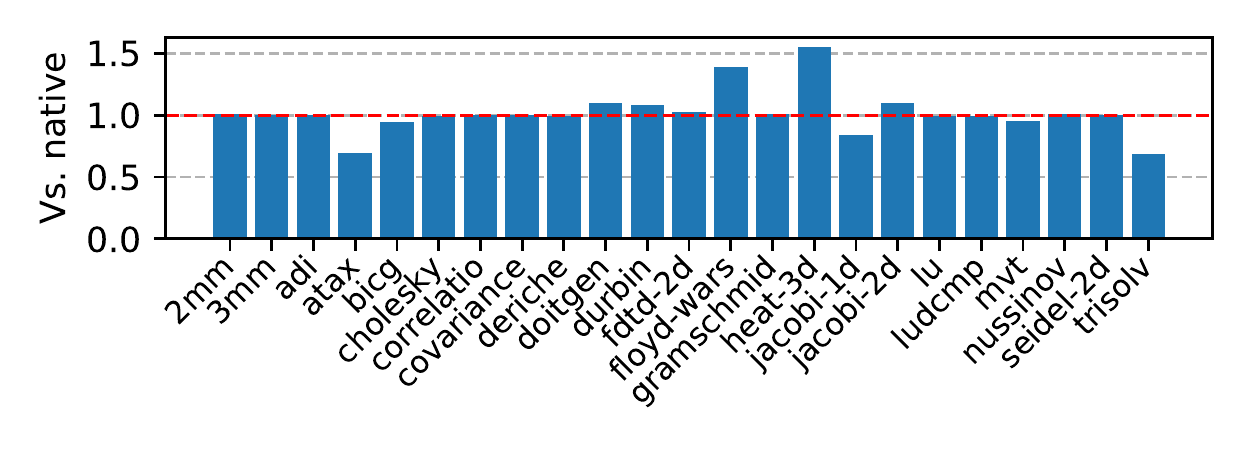}                    
      \caption{Polybench}
      \label{fig:Polybench}        
  \end{subfigure}  
  \begin{subfigure}[]{\linewidth}
      \includegraphics[width=\linewidth]{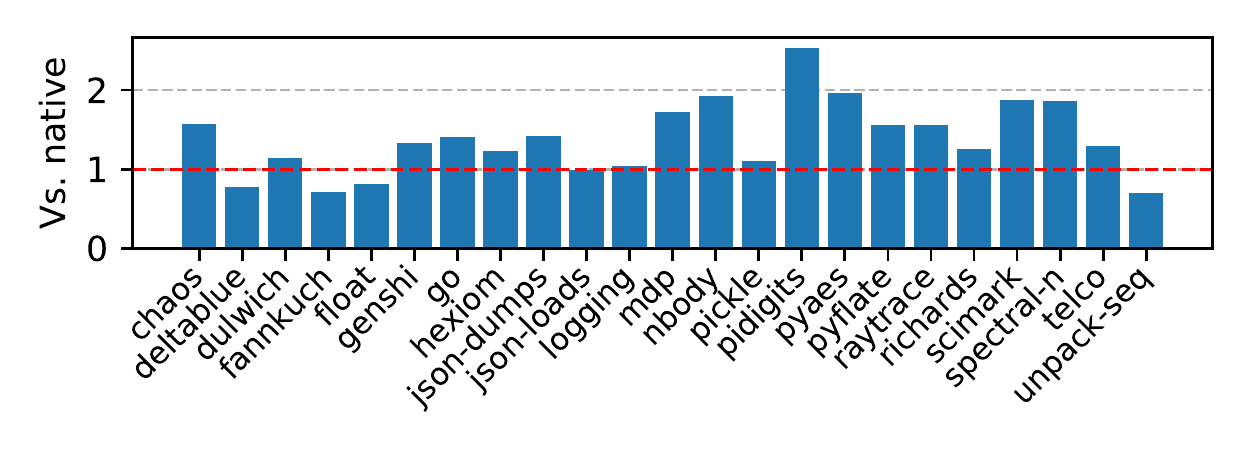}                    
      \caption{Python Performance Benchmark}
      \label{fig:PythonBench}        
  \end{subfigure}    
  \caption{Performance of Faaslets with Python}
  \label{fig:MicrobenchResults}
\end{figure}

\F\ref{fig:MicrobenchResults} shows the performance overhead when running 
both sets of benchmarks compared to native execution. All but two of the 
Polybench benchmarks are comparable to native with some showing performance 
gains. Two experience a 40\%--55\% overhead, both of which benefit 
from loop optimisations that are lost through compilation to WebAssembly.
Although many of the Python benchmarks are within a 25\% overhead or better, 
some see a 50\%--60\% overhead, with \code{pidigits} showing a 240\% overhead. 
\code{pidigits} stresses big integer arithmetic, which incurs significant 
overhead in 32-bit WebAssembly.

% Interpretation
Jangda~\etal\cite{NotSoFast:2019} report that code compiled to WebAssembly has
more instructions, branches and cache misses, and that these overheads are compounded
on larger applications. Serverless functions, however, typically are not complex
applications and operate in a distributed setting in which distribution
overheads dominate. As shown in \F\ref{fig:MatrixDuration}, \sys can achieve
competitive performance with native execution, even for functions
interpreted by a dynamic language runtime. 

\subsection{Efficiency of Faaslets vs. Containers} \label{sec:eval:overheads}

\begin{table}[t]
  \centering\footnotesize
  \begin{tabular}{lrrrr}
    \toprule
    & \textbf{Docker} & \textbf{Faaslets} & \textbf{Proto-Faaslets} & \textbf{vs. Docker} \\    
    \midrule
    Initialisation & 2.8\unit{s}  & 5.2\unit{ms} & 0.5\unit{ms}  & \textbf{5.6K$\times$} \\
    CPU cycles & 251M  & 1.4K & 650 & \textbf{385K$\times$} \\
    PSS memory & 1.3\unit{MB}  & 200\unit{KB} & 90\unit{KB} & \textbf{15$\times$} \\
    RSS memory & 5.0\unit{MB}  & 200\unit{KB} & 90\unit{KB} & \textbf{57$\times$} \\
    Capacity & \textasciitilde{}8\unit{K} & \textasciitilde{}70\unit{K} & >100\unit{K} & \textbf{12$\times$} \\
    \bottomrule
  \end{tabular}
  \caption{Comparison of Faaslets vs. container cold starts\\ (no-op function)}
  \label{table:Efficiency}
\end{table}

% Goal 
Finally we focus on the difference in footprint and cold-start initialisation latency between Faaslets and 
containers.

% Set up
To measure memory usage, we deploy increasing numbers of parallel functions 
on a host and measure the change in footprint with each extra function. 
Containers are built from the same minimal image (\code{alpine:3.10.1}) so can access 
the same local copies of shared libraries. To highlight the impact of this 
sharing, we include the proportional set size~(PSS) and resident set size~(RSS) 
memory consumption. Initialisation times and CPU cycles are measured across 
repeated executions of a no-op function. We observe the capacity as the maximum 
number of concurrent running containers or Faaslets that a host can sustain 
before running out of memory.

% Results
\T\ref{table:Efficiency} shows several orders of magnitude improvement in CPU
cycles and time elapsed when isolating a no-op with Faaslets, and a
further order of magnitude using Proto-Faaslets. With an optimistic PSS memory
measurement for containers, memory footprints are almost seven times lower using
Faaslets, and 15$\times$ lower using Proto-Faaslets. A single host
can support up to 10$\times$ more Faaslets than containers, growing to
twelve times more using Proto-Faaslets. 

To assess the impact of restoring a non-trivial Proto-Faaslet snapshot,
we run the same initialisation time measurement for a Python no-op function. The
Proto-Faaslet snapshot is a pre-initialised CPython interpreter, and the
container uses a minimal \code{python:3.7-alpine} image. The container initialises in
3.2\unit{s} and the Proto-Faaslet restores in 0.9\unit{ms}, demonstrating a
similar improvement of several orders of magnitude.

\begin{figure}[t]
  \centering
  \includegraphics[width=.95\linewidth]{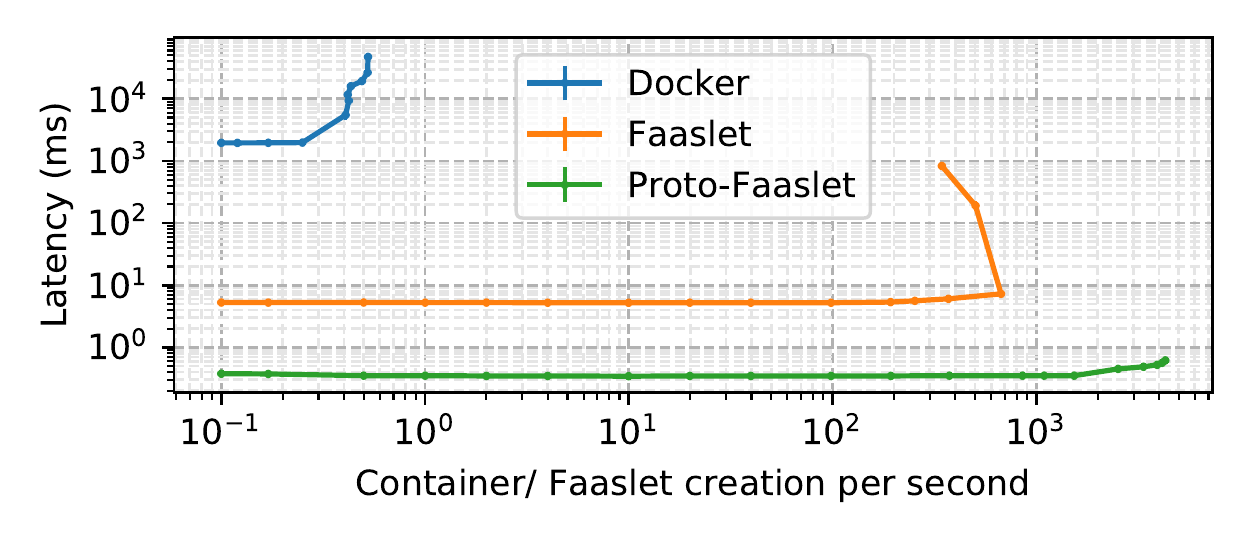}
  \caption{Function churn for Faaslets vs. containers}    
  \label{fig:MicroTptLat}
\end{figure}

To further investigate cold-start initialisation times, we measure the
time to create a new container/Faaslet at increasingly higher rates of
cold-starts per second. We also measure this time when restoring the
Faaslet from a Proto-Faaslet. The experiment executes on a single host, with
the containers using the same minimal image.

\F\ref{fig:MicroTptLat} shows that both Faaslets and containers maintain a
steady initialisation latency at throughputs below 3\unit{execution/s}, with
Docker containers initialising in \textasciitilde{}2\unit{s} and Faaslets in
\textasciitilde{}5\unit{ms} (or \textasciitilde{}0.5\unit{ms}
when restored from a Proto-Faaslet). As we increase the churn in Docker past
3\unit{execution/s}, initialisation times begin to increase with no gain in
throughput. A similar limit for Faaslets is reached at around
600\unit{execution/s}, which grows to around 4000\unit{execution/s}
with Proto-Faaslets.

% Interpretation
We conclude that Faaslets offer a more efficient and performant form of
serverless isolation than Docker containers, which is further improved
with Proto-Faaslets. The lower resource footprint and initialisation times of
Faaslets are important in a serverless context. Lower resource footprints reduce
costs for the cloud provider and allow a higher packing density of parallel
functions on a given host. Low initialisation times reduce cost and latency for
the user, through their mitigation of the cold-start problem.

% End

%% file: sections/related.tex
% Related work

\section{Related Work} \label{sec:rel_work}

\mypar{Isolation mechanisms} Shreds~\cite{Shreds:2016} and
Wedge~\cite{Wedge:2008} introduce new OS-level primitives for memory isolation,
but focus on intra-process isolation rather than a complete executable as
Faaslets do. Light-weight Contexts~\cite{LightweightContexts:2016} and
Picoprocesses~\cite{Picoprocesses:2013} offer lightweight sandboxing of
complete POSIX applications, but do not offer efficient shared state.

\mypar{Common runtimes} Truffle~\cite{Truffle:2013} and
GraalVM~\cite{GraalVM:2013} are runtimes for language-independent bytecode; the
JVM also executes multiple languages compiled to Java
bytecode~\cite{chiba2003easy}. Despite compelling multi-language support, none
offer multi-tenancy or resource isolation. GraalVM has recently added support
for WebAssembly and could be adapted for Faaslets.

\mypar{Autoscaling storage} \sys's global state tier is currently implemented
with a distributed Redis instance scaled by Kubernetes horizontal pod
autoscaler~\cite{KubernetesScalingWebsite}. Although this has not been a
bottleneck, better alternatives exist: Anna~\cite{Anna:2019} is a distributed
KVS that achieves lower latency and more granular autoscaling than Redis;
Tuba~\cite{Tuba:2014} provides an autoscaling KVS that operates within
application-defined constraints; and Pocket~\cite{Pocket:2018} is a granular
autoscaled storage system built specifically for a serverless
environments. Crucial~\cite{Crucial:2019} uses
Infinispan~\cite{Infinispan:2012} to build its distributed object storage,
which could also be used to implement \sys's global state tier.

\noindent\mypar{Distributed shared memory~(DSM)} FaRM~\cite{Farm:2014} and
RAMCloud~\cite{Ramcloud:2015} demonstrate that fast networks can overcome
the historically poor performance of DSM
systems~\cite{MuninImplementation:1991}, while DAL~\cite{DAL:2017} demonstrates
the benefits of introducing locality awareness to DSM. \sys{}'s global tier could
be replaced with DSM to form a distributed object store, which would require a
suitable consensus protocol, such as Raft~\cite{Raft:2014}, and a communication
layer, such as Apache Arrow~\cite{ApacheArrowWebsite}.

\mypar{State in distributed dataflows} Spark~\cite{RDD:2012} and
Hadoop~\cite{Hadoop:2010} support stateful distributed computation. Although
focuses on fixed-size clusters and not fine-grained elastic scaling or
multi-tenancy, distributed dataflow systems such as Naiad~\cite{Naiad:2013},
SDGs~\cite{StatefulDataflow:2014} and CIEL~\cite{CIEL:2011} provide high-level
interfaces for distributed state, with similar aims to those of distributed
data objects---they could be implemented in or ported to
\sys. Bloom~\cite{Bloom:2011} provides a high-level distributed programming
language, focused particularly on flexible consistency and replication, ideas
also relevant to \sys.

\mypar{Actor frameworks} Actor-based systems such as
Orleans~\cite{Orleans:2011}, Akka~\cite{AkkaWebsite} and Ray~\cite{Ray:2018}
support distributed stateful tasks, freeing users from scheduling and state
management, much like \sys. However, they enforce a strict asynchronous
programming model and are tied to a specific languages or language runtimes,
without multi-tenancy.

% End

%% file: sections/conclusion.tex
% Conclusions

\section{Conclusions}
\label{sec:concl}

To meet the increasing demand for serverless big data, we presented \sys, a
runtime that delivers high-performance efficient state without compromising
isolation. \sys executes functions inside Faaslets, which provide memory safety
and resource fairness, yet can share in-memory state. Faaslets are initialised
quickly thanks to Proto-Faaslet snapshots. Users build stateful serverless
applications with distributed data objects on top of the Faaslet state
API. \sys's two-tier state architecture co-locates functions with required
state, providing parallel in-memory processing yet scaling across hosts. The
Faaslet host interface also supports dynamic language runtimes and 
traditional POSIX applications.

%  End

%% file: sections/acks.tex
\label{sec:acks}

\noindent\mypar{Acknowledgements} We thank the anonymous reviewers and our shepherd, Don
Porter, for their valuable feedback. This work was partially supported by the
European Union’s Horizon 2020 Framework Programme under grant agreement
825184~(CloudButton), the UK Engineering and Physical Sciences Research
Council~(EPSRC) award 1973141, and a gift from Intel Corporation under the TFaaS
project.

% End